\documentclass[
    ,final            % use final for the camera ready runs
  ]
  {aipproc}

\layoutstyle{8x11single}

\begin{document}

\title{The Y(3940), Z(3930) and the X(4160) as dynamically generated resonances from the vector-vector interaction}


\author{R. Molina}{
  address={Departamento de F\'{\i}sica Te\'orica and IFIC,
Centro Mixto Universidad de Valencia-CSIC,
Institutos de Investigaci\'on de Paterna, Aptdo. 22085, 46071 Valencia,
 Spain}
}

\author{E. Oset}{
  address={Departamento de F\'{\i}sica Te\'orica and IFIC,
Centro Mixto Universidad de Valencia-CSIC,
Institutos de Investigaci\'on de Paterna, Aptdo. 22085, 46071 Valencia,
 Spain}
  } % additional visiting address

\begin{abstract}
 We study the vector-vector interaction within the framework of the hidden gauge formalism for the sector with quantum numbers charm $C=0$ and strangeness $S=0$ in the region around $4000$ MeV. We get five poles, three of which could be identified with the Y(3940), Z(3930) and X(4160). These poles appear with quantum numbers $I=0$ and $J^{PC}=0^{++},2^{++}$ and $2^{++}$, respectively, and can be considered as hadronic molecules made of $D^*\bar{D}^*$, $D_s^*\bar{D}^*_s$.
\end{abstract}

\maketitle

%%%%%%%%%%%%%%%%%%%%%%%%%%%%%%%%%%%%%%%%%%%%
%% MAINMATTER
%%%%%%%%%%%%%%%%%%%%%%%%%%%%%%%%%%%%%%%%%%%%

\section{Introduction}

 Recently, the B-factories at SLAC, KEK and CESR, which were originally constructed to test matter-antimatter asymmetries or CP violation, have discovered new hidden-charm states around the energy region of $4000$ MeV. These new states do not seem to have a simple $c\bar{c}$ structure. They are naively called as XYZ particles. Some of them, which we consider in this manuscript, are the X(3940), the Y(3940), the X(4160) and the Z(3930). 

The X(3940) and the X(4160) were observed by the Belle Collaboration in the $e^+e^-\to J/\psi X$ reaction as a $D\bar{D}^*$ and $D^*\bar{D}^*$ mass peak respectively \cite{Abe2}. The $X(3940)\to D\bar{D}^*$ has been observed but there is no signal for the $D\bar{D}$ or the $\omega J/\psi$ decays. Because the fact that the $\eta_c(1S)$ and $\eta_c(2S)$ were also produced in double-charm production, it was believed that the X(3940) could have $J^{PC}=0^{-+}$, being a $3^1 S_1$ charmonium state ($\eta''_c$), but then the X(3940) should have a mass $\sim$ $4050$ MeV or even higher \cite{Barnes}. Concerning to the X(4160), the known charmonium states seen from a $e^+e^-\to J/\psi D^*\bar{D^*}$ reaction have $J=0$, thus, if this state is identified with a $3^1S_0$ ($\eta''_c$) or $4^1S_0$ ($\eta'''_c$) charmonium state, the mass predicted in the first case will be smaller ($4050$ MeV) and higher ($4400$ MeV) in the second case \cite{Barnes}. On the other hand, the Y(3940) has been firstly observed by the Belle Collaboration and Babar has confirmed it in $B\to K Y\to K\omega J/\psi$ decays \cite{Abe4}. Nevertheless, the values for the mass and width reported by Babar are smaller than the Belle's values. Also, very recently, Belle has reported a narrow peak in the cross section for $\gamma\gamma\to \omega J/\psi$ \cite{olsen2} that is consistent with the mass and width reported for the Y(3940) by the Babar group. This peak, dubbed as X(3915), has a mass $M=3914\pm 4\pm 2$ MeV and width $\Gamma=28\pm12^{+2}_{-8}$ MeV. The measure of the mass favor the possible assignment of this new peak to the Y(3940) instead of the Z(3930). The production modes ensure that the X(4160) as well as the X(3940) and Y(3940) have C-parity positive.
 
 Recently, the CDF Collaboration at Fermilab has announced a narrow peak near the $J/\psi \phi$ threshold, which is designated as Y(4140), observed in the $B^+\to J/\psi \phi K^+$ decay, it has a mass $M=4143\pm 2.9 (stat) \pm 1.2 (syst)$ MeV and a width $\Gamma=11.7^{+8.3}_{-5.0} (stat) \pm 3.7 (syst)$ MeV \cite{CDF}. The width observed is quite different from the one reported by Belle for the X(4160) \cite{Abe3}, suggesting that one is taking about a different state.
     
Finally, the Z(3930) has been seen by Belle as a peak in the spectrum of $D\bar{D}$ mesons produced in $\gamma\gamma$ collisions. The Belle measurements favors the $2^{++}$ hypothesis, making the assignment of the Z(3930) to the $2^3 P_2(\chi'_{c2})$ charmonium state possible \cite{Godfrey,Chao}.

In \cite{Liu, Tania, Bracco}, previous theoretical work about the origin of these particles has been done, supporting the interpretation of the Y(3940) and the Y(4140) as $D^*\bar{D}^*$ and $D_s^*\bar{D}_s^*$ molecules respectively. 

In this paper we will propose a theoretical explanation on the nature of some of these XYZ states, providing structure and quantum numbers for them. Our work is based on the hidden gauge symmetry (HGS) formalism for the interaction of vector mesons, which was introduced by Bando-Kugo-Yamawaki \cite{hidden}. Concretely, we study the case of a system of two vector meson with charm $C=0$ and strangeness $S=0$ around $4000$ MeV (hidden charm sector).

\section{Formalism: The $VV$ interaction}
\begin{figure}[htb]
\centering
\includegraphics*[width=8cm]{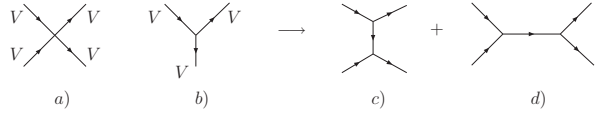}
\caption{Terms of the ${\cal L}_{III}$ Lagrangian: a) four vector contact term,
 Eq.~(\ref{lcont}); b) three-vector interaction, Eq.~(\ref{l3V}); c) $t$ and 
 $u$ channels from vector exchange; d) $s$ channel for vector exchange.}
\label{fig:3V4V}
\end{figure}

Within the theorical framework, there are two main ingredients: first, we take the Lagrangians for the interaction of vector mesons among themselves, that come from the hidden gauge formalism of Bando-Kugo-Yamawaki \cite{hidden}. Second, we introduce the potential $V$ obtained from these Lagrangians (projected in s-wave, spin and isospin) in the Bethe Salpeter equation:
\begin{equation}
T= (\hat{1}-VG)^{-1} V\ ,
\label{Bethe}
\end{equation}
where $G$ is the loop function. Therefore, we are summing all the diagrams containing zero, one, two... loops implicit in the Bethe Salpeter equation. Finally, we look for poles of the unitary $T$ matrix in the second Riemann sheet. All this procedure is well explained in \cite{raquel,geng}. In these works, two main vertices are taken into account for the computation of the potential $V$: the four-vector-contact term and the three-vector vertex, which are provided respectively from the Lagrangians:
\begin{equation}
{\cal L}^{(c)}_{III}=\frac{g^2}{2}\langle V_\mu V_\nu V^\mu V^\nu-V_\nu V_\mu
V^\mu V^\nu\rangle\ ,
\label{lcont}
\end{equation}
\begin{equation}
{\cal L}^{(3V)}_{III}=ig\langle (\partial_\mu V_\nu -\partial_\nu V_\mu) V^\mu V^\nu\rangle
\label{l3V}\ .
\end{equation}
By means Eq. (\ref{l3V}), the vector exchange diagrams in Fig. \ref{fig:3V4V} are calculated. The diagram in Fig. \ref{fig:3V4V}d) leads to a repulsive p-wave interaction for equal masses of the vectors \cite{raquel} and only to a minor component of s-wave in the case of different masses \cite{geng}. Thus, the four-vector-contact term in Fig. \ref{fig:3V4V}a) and the t(u)-channel vector exchange diagrams in Fig. \ref{fig:3V4V}c) are responsible for the generation of resonances or bound states if the interaction is strong enough. 

As we are interested in the region close to the two vector meson threshold, the three momenta of the external particles can be neglected compared with the mass of the vector meson, and thus, we can make $|\vec{q}|/M\sim 0$ for external particles, which considerably simplify the calculation.

\begin{figure}[htb]
\centering
\includegraphics*[width=8cm]{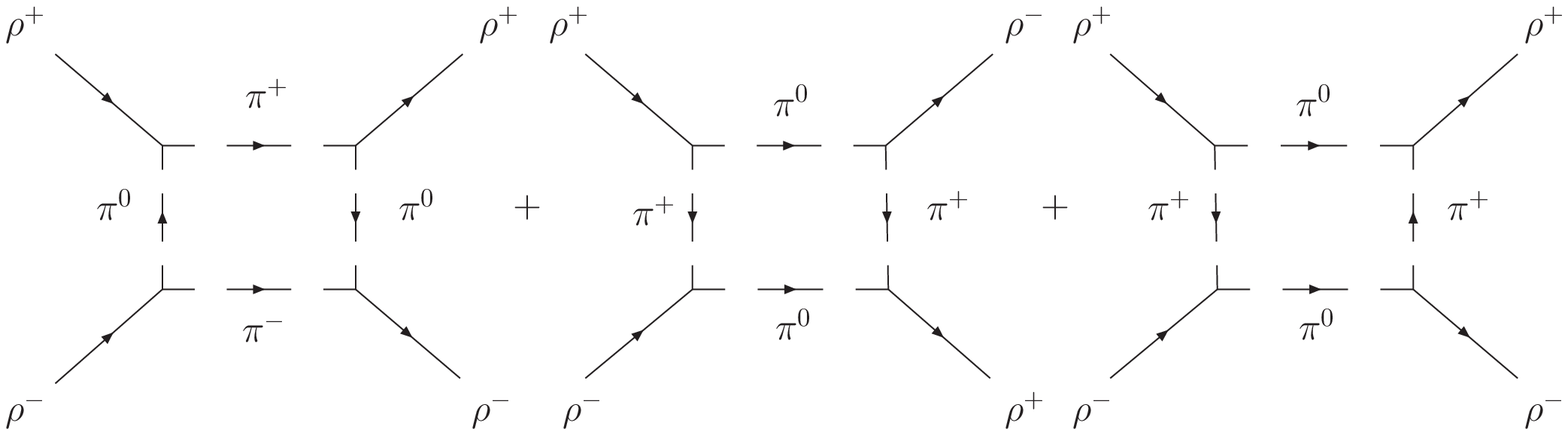}
\caption{Box diagrams for the case of the $\rho\rho$ interaction.}
\label{fig:box}
\end{figure}

In order to consider the pseudoscalar-pseudoscalar decay mode, and thus compare properly with the available experimental data for the width, a box diagram is also included, which we show in Fig. \ref{fig:box} for the particular case of the $\rho\rho$ system. Crossed box diagrams and box diagrams involving anomalous couplings were also calculated in \cite{raquel}, but they were much smaller, specially in the case of the anomalous coupling, than the contributions coming from the box diagrams in Fig. \ref{fig:box}, and they were not considered in later works \cite{geng,raquel2,raquel3}. In \cite{raquel}, two states were obtained within this formalism for the particular case of the $\rho \rho$ system that were identified with the $f_2(1270)$ and the $f_0(1370)$ as $\rho\rho$ molecular states. Whereas in \cite{geng}, the procedure of \cite{raquel} is extended to SU(3) and eleven states are found, five of them can be associated with data in the PDG: the $f_0(1370)$, $f_0(1710)$, $f_2(1270)$, $f_2'(1525)$ and $K^*_2(1430)$. In these works, the integral of two-vector-meson loop function can be calculated by means of the dimensional regularization method or by means of the use of a cutoff. This requires the introduction of one parameter, which is the subtraction constant ( or a cutoff parameter), that is tuned to reproduce the mass of the tensor states, whereas the other states are predicted. The box diagrams need two parameters, the cutoff for the integral, $\Lambda$, and one parameter, $\Lambda_b$, involved in the form factor used. These parameters were considered to be around $1$ GeV and $1.4$ GeV respectively in \cite{raquel} to get reasonable values of the $f_0(1370)$ and $f_2(1270)$ widths. In the later work of \cite{geng}, these values also provide a good description of the widths for the other states.

In \cite{raquel2}, the work of \cite{raquel, geng} is extended to SU(4) in the study of the $\rho(\omega)D^*$ system.
Here, the $V$ matrix is straigthforward extended to SU(4), as it was done in the work of \cite{Gamermann}:
\begin{equation}
V_\mu=\left(
\begin{array}{cccc}
\frac{\rho^0}{\sqrt{2}}+\frac{\omega}{\sqrt{2}}&\rho^+& K^{*+}&\bar{D}^{*0}\\
\rho^-& -\frac{\rho^0}{\sqrt{2}}+\frac{\omega}{\sqrt{2}}&K^{*0}&D^{*-}\\
K^{*-}& \bar{K}^{*0}&\phi&D^{*-}_s\\
D^{*0}&D^{*+}&D^{*+}_s&J/\psi\\
\end{array}
\right)_\mu \ ,
\label{Vmu}
\end{equation}
where the ideal mixing has been taken for $\omega$, $\phi$ and $J/\psi$. The SU(4) symmetry is strongly broken by the explicit use of the masses of the vector mesons in the vector-exchange diagrams, where the terms of the potential that comes from the diagrams that exchange a $D^*$ meson are proportional to $\kappa=\frac{m^2_\rho}{m^2_{D^*}}\sim 0.15$, and this gives rise to corrections of the order of $10$\% of the $\rho$-exchange terms. In this work, three states are obtained with $I=1/2$, $J=0, 1$ and $2$ respectively, whereas the interaction is repulsive for $I=3/2$ and thus no exotics states appear. The subtraction constant $\alpha$ in the function loop is fine-tuned around its natural value of $-2$ \cite{oller,gamerman2}, in order to get the position of the $D^*_2(2460)$ state at the PDG.  The interesting thing comes from the calculation of the box diagram, this box diagram only has $J=0$ and $2$, therefore, the state found with $J=1$ cannot decay to $\pi D$ by means of this mechanism, which is the reason why we get a small width  of $4$ MeV, compared with that of the state for $J=2$, around $40$ MeV \cite{raquel2}. Thus, we find a reasonable 
explanation for why the $D^*(2640)$ has a small width, $\Gamma<15$ MeV, when we associate to this state the quantum numbers $J^{P}=1^{+}$. The state for $J=0$ with mass around $2600$ MeV is a prediction of the model and there is no data over this state up to now.

As one can see, the model in SU(3) provides a good description of the properties of many physical states. Also, in SU(4) we have seen that interesting new states appear. Thus, in the next section, we study the case of $C=0$ and $S=0$ around the region of $4000$ MeV.

\section{The XYZ particles}
In view of the results obtained in the cases of the vector-vector interaction in SU(3) and the $\rho$($\omega$) $D^*$ system in \cite{raquel2} we make an attempt to study now the sector with $C=0$ and $S=0$. In the case of $I=0$ we have 10 channels: $D^*\bar{D}^*(4017)$, $D^*_s\bar{D}^*_s(4225)$, $K^*\bar{K}^*(1783)$, $\rho\rho(1551)$, $\omega\omega(1565)$, $\phi\phi(2039)$, $J/\psi J/\psi (6194)$, $\omega J/\psi (3880)$, $\phi J/\psi (4116)$ and $\omega \phi (1802)$, and 6 channels for $I=1$: $D^*\bar{D}^*(4017)$, $K^*\bar{K}^*(1783)$, $\rho\rho(1551)$, $\rho\omega(1558)$, $\rho J\psi(3872)$ and $\rho\phi (1795)$. The procedure is identical to the one explained in the previou section, see \cite{raquel3} for practical details, except by the complication added due to the many channels involved. The function loop is computed by means of dimensional regularization and we fix the value of $\mu=1000$ MeV as in the work of \cite{geng}. We establish two different subtraction constants, one for the channels involving two light vector meson, which is called as $\alpha_L$, and the other one, for the channels involving two heavy vector meson: $\alpha_H$. When one channel involves both kind of mesons, one heavy and one light vector meson,  we also put $\alpha_L$. The value of $\alpha_L$ is set to get the position of the pole $f_2(1270)$ as $\alpha_L=-1.65$, like it was done in the work of \cite{geng}. Thus, we get the correct position of the poles found in this work. The $\alpha_H$ constant is set to $-2.07$ in order to get the position of the pole found in $S=0$ around $3940$ MeV. By the fact of having several channels, ones involving heavy mesons and the other light mesons, the $g$ parameter in the Lagrangian must be different for the different cases. Thus, we put $g=M_\rho/(2 f_\pi)=4.17$ for light mesons, and $g_D=M_{D^*}/(2 f_{D})=6.9$, $g_{D_s}=M_{D^*_s}/(2 f_{D_s})=5.47$, $g_{\eta_c}= M_{J/\psi}/(2 f_{\eta_c})=5.2$, when one $D^*$, $D_s^*$ or $J/\psi$ meson is involved. The $D\bar{D}$ decay mode is also considered by means of a box diagram as it was done in \cite{raquel2}, but the effect is small compared with the (light) vector-(light or heavy) vector decay modes. The results for the pole position obtained together with a possible assignment with some of the XYZ particles observed in the energy region of $4000$ MeV is given in Table \ref{tab:thexp}. In order to do a proper comparison with the experiment, we give in Table \ref{tab:olsen} the XYZ discovered up to now in which we are interested, which is taken from \cite{Godfrey}.

As it is shown in Table \ref{tab:thexp}, we find one pole with a mass $\sim 3940$ MeV for each spin $J=0,1$ and $2$ and $I=0$. The modules of the coupling contants to the different channels are obtained from the residues of the amplitudes and they are given in Tables \ref{tab:res0} and \ref{tab:res1}. From Table \ref{tab:res0}, we see that these three poles around $3940$ MeV are about $50-70$\% $D^*\bar{D}^*$ and $20-30$\% $D^*_s\bar{D}^*_s$, whereas all the other channels account only for $10-30$\% or less. In Table \ref{tab:olsen} (experimental data) there are also three states around this mass. The state that we found for $I=1$ is discarded to be associated with one of these states, because in our model this state decays to $D\bar{D}$ and not to $\omega J/\psi$, thus cannot be associated with the X(3940), since the $D\bar{D}$ decay has not been observed \cite{Abe2}, whereas it cannot be associated to the Y(3940) either, because the experimental state decays to $\omega J/\psi$. On the other hand, we find a width so large that it is likely to be assigned to the Z(3930). In that way, the state found for $I=1$ is a prediction of the model. In Table \ref{tab:thexp} we call it as $Y_p(3912)$, where '$p$' stands for prediction. For this state, we find that the $D^*\bar{D}^*$ channel contributes $60$\%, whereas the sum of the modules of the couplings for the rest of the channels gives $\sim 40$\%.

Although, at first one could think that the three states that we found around $3940$ MeV for $I=0$ could correspond to those in Table \ref{tab:olsen}, this assignment can lead to error. The reason is that all the three experimental states have C-parity positive, while our state for $J=1$ has C-parity negative. From the experimental point of view, this is clear by the production mechanism: $e^+e^-$ has C-parity negative (it comes from a photon) and as the $J/\psi$ has also C-parity negative, the X(3940) must have C-parity positive. Thus, the state that we found for $J=1$ is also a prediction.

 In Table \ref{tab:olsen}, we find that there are two different experimental measures for the mass and width of the Y(3940). One is done by Babar and the other one by Belle with bigger values for both magnitudes. The point in favor to the assignment of the state that we found with $J^{PC}=0^{++}$ to the Y(3940) is the calculation of the $\Gamma((3943,0^+[0^{++}])\to \omega J/\psi)$ that can be done straightforwardly by means of the formula:
\begin{equation}
\Gamma((3943,0^{+}[0^{++}])\to \omega J/\psi)=\frac{p\,|g_{Y\omega J/\psi}|^2}{8 \pi\,M_Y^2} 
\label{decay}
\end{equation}
where $p$ is the momentum of $\omega$ in the resonance rest frame. Taking the modules of the coupling $|g_{Y\omega J/\psi}|=1445$ MeV from Table \ref{tab:res0}, we obtain $\Gamma((3943,0^+[0^{++}])\to \omega J/\psi)=1.52$ MeV, compatible with the expected experimental value for this decay, $\Gamma(Y(3940)\to \omega J/\psi)>1$ MeV \cite{Godfrey}. Then, we have found a natural explanation on why this rate is much larger than it would be if it corresponded to a hadronic transition between charmonium states. 

On the other hand, the two states that we find in $I=0$ for $J=2$, with masses $M=3922$ and $4157$ MeV, are assigned to the Z(3930) and X(4160) respectively, by the proximity of the mass, width and quantum numbers. The structure of the state found with $M=4157$ MeV is radically different to the other states. We see from Table \ref{tab:res0}, that the $D^*_s\bar{D}^*_s$ coupling accounts for the $49$\%, whereas the $D^*\bar{D}^*$ coupling gives only the $3$\%, being the sum of the contributions of all the other channels of $\sim48$\%.
\begin{table}
\begin{tabular}{lrrrrr}
\hline
 \tablehead{1}{l}{b}{$I^G[J^{PC}]$}
  & \tablehead{1}{r}{b}{Theory}
  & \tablehead{4}{c}{b}{Experiment}
  \\
\hline
& (Mass, Width) & Name & Mass & Width &$J^{PC}$\\
$0^+[0^{++}]$&($3943$, $17$)&$Y(3940)$&$3943\pm 17$&$87\pm 34$&$J^{P+}$\\
& & & $3914.3^{+4.1}_{-3.8}$ & $33^{+12}_{-8}$ & \\
& & & $3914\pm 4\pm 2$ & $28\pm 12^{+2}_{-8}$ & \\
$0^-[1^{+-}]$&($3945$, $0$)&"$Y_p(3945)$"& & & \\
$0^+[2^{++}]$&($3922$, $55$)&$Z(3930)$&$3929\pm 5$& $29\pm 10$&$2^{++}$\\
$0^+[2^{++}]$&($4157$, $102$)&$X(4160)$&$4156\pm 29$& $139^{+113}_{-65}$& $J^{P+}$\\
\hline
$1^-[2^{++}]$&($3912$, $120$)&"$Y_p(3912)$"& & & \\
\hline
\end{tabular}
\caption{Comparison of the mass, width and quantum numbers with the experiment. All the quantities are in units of MeV.}
\label{tab:thexp}
\end{table}

\begin{table}
\begin{tabular}{lrrrrr}
\hline
 \tablehead{1}{l}{b}{State}
  & \tablehead{1}{r}{b}{M (MeV)}
  & \tablehead{1}{r}{b}{$\Gamma$ (MeV)}
  & \tablehead{1}{r}{b}{$J^{PC}$}
  & \tablehead{1}{r}{b}{Decay modes}
  &  \tablehead{1}{r}{b}{Production modes}
 \\
\hline
$Z(3930)$& $3929\pm5$ & $29\pm 10$ & $2^{++}$ & $D\bar{D}$ & $\gamma\gamma$ \\
$X(3940)$ & $3942\pm 9$ &$37\pm 17$ & $J^{P+}$ & $D\bar{D}^*$ & $e^+ e^-\to J/\psi X(3940)$\\
$Y(3940)$ & $3943\pm 17$  & $87\pm 34$& $J^{P+}$ & $\omega J/\psi$ & $B\to KY(3940)$\\
 & $3914.3^{+4.1}_{-3.8}$& $33^{+12}_{-8}$& & & \\
$X(4160)$ & $4156\pm 29$ & $139^{+113}_{-65}$& $J^{P+}$ & $D^{*}\bar{D}^*$&$e^+e^-\to J/\psi X(4160)$\\
\hline
\end{tabular}
\caption{Properties of the candidate XYZ mesons.}
\label{tab:olsen}
\end{table}

\begin{table}
\begin{tabular}{lr|rrrrrrrrrr}
\hline
 \tablehead{1}{l}{b}{$I^G[J^{PC}]$}
  & \tablehead{1}{r}{b}{$\sqrt{s}_{pole}$}
  & \tablehead{1}{r}{b}{$D^*\bar{D}^*$}
  & \tablehead{1}{r}{b}{$D^*_s\bar{D}_s^*$}
  & \tablehead{1}{r}{b}{$K^*\bar{K}^*$}
  &  \tablehead{1}{r}{b}{$\rho\rho$}
  & \tablehead{1}{r}{b}{$\omega\omega$}
  & \tablehead{1}{r}{b}{$\phi\phi$}
  & \tablehead{1}{r}{b}{$J/\psi J/\psi$}
  & \tablehead{1}{r}{b}{$\omega J/\psi$}
  &  \tablehead{1}{r}{b}{$\phi J/\psi$}
  &  \tablehead{1}{r}{b}{$\omega\phi$}
\\
\hline
 $0^+[0^{++}]$&$3943 + i 7.4$&$18822$&$8645$&$15$&$52$&$1368$&$1011$&$422$&$1445$&$910$&$240$\\
$0^-[1^{+-}]$&$3945 +i 0$&$18489$&$8763$&$40$&$0$&$0$&$0$&$0$&$0$&$0$&$0$\\
$0^+[2^{++}]$&$3922+i 26$&$21177$&$6990$&$44$&$84$&$2397$&$1999$&$1794$&$2433$&$3061$&$789$\\
$0^+[2^{++}]$&$4169+i 66$&$1319$&$19717$&$87$&$73$&$2441$&$3130$&$2841$&$2885$&$5778$&$1828$\\
\hline
\end{tabular}\\
\caption{Quantum numbers, pole positions and modules of the couplings $|g_{i}|$ in units of MeV for $I=0$.}
\label{tab:res0}
\end{table}

\begin{table}
\begin{tabular}{lr|rrrrrr}
\hline
 \tablehead{1}{l}{b}{$I^G[J^{PC}]$}
  & \tablehead{1}{r}{b}{$\sqrt{s}_{pole}$}
  & \tablehead{1}{r}{b}{$D^*\bar{D}^*$}
  & \tablehead{1}{r}{b}{$K^*\bar{K}^*$}
  &  \tablehead{1}{r}{b}{$\rho\rho$}
  & \tablehead{1}{r}{b}{$\rho\omega$}
  & \tablehead{1}{r}{b}{$\rho J/\psi$}
  & \tablehead{1}{r}{b}{$\rho \phi$}
\\
\hline
$1^-[2^{++}]$&$3919+ i74$&$20869$&$152$&$0$&$3656$&$6338$&$2731$\\
\hline
\end{tabular}
\caption{Quantum numbers, pole position and modules of the couplings $|g_{i}|$ in units of MeV for $I=1$.}
\label{tab:res1}
\end{table} 
\section{Conclusions}
We have studied the $C=0$ and $S=0$ sector around the energy region of $4000$ MeV within the hidden gauge formalism combined with coupled channel unitarity. We find one state with mass $\sim 3940$ MeV for each spin $J=0,1$ and $2$. Only the states with $J=0$ and $2$ can be associated the XYZ particles with mass $\sim 3940$ MeV in the PDG. Concretely, we associate the $J^{PC}=0^{++}$ state with the Y(3940), and the $2^{++}$ state with Z(3930). Nevertheless, the $J^{PC}=1^{+-}$ cannot be associated with known states and it becomes a prediction. The only state that we find for $I=1$,  with $J^{PC}=2^{++}$, cannot be associated either, because of the decay modes and the large width. This new prediction could be observed in the $\rho J/\psi$ channel. For those states with masses around $3940$ MeV, we find that the $D^*\bar{D}^*$ channel accounts for the $50-70$\%. 
Our model predicts another state with mass around $4160$ MeV and quantum numbers $I^G[J^{PC}]=0^+[2^{++}]$, which we identify as the X(4160) by the proximity of the mass, width and C-parity. This state has a different structure, being the $D*_s\bar{D}^*_s$ channel the one contributing most to the generation of the state. 

The region around $3940$ MeV is very interesting and there could be more resonances not yet seen in this region. The findings of this work should motivate the experimentalist to look into this region in the channels that involve light vector - light vector or light vector - heavy vector like $K^*\bar{K}^*$ and $\rho J/\psi$.
\bibliographystyle{aipproc}

\end{document}